\documentstyle[11pt,newpasp,twoside,epsf]{article}
\def\edcomment#1{\iffalse\marginpar{\raggedright\sl#1\/}\else\relax\fi}
\reversemarginpar

\begin{document}
\title{RR Lyrae variables in $\omega$ Centauri: a theoretical route}
\author{V. Castellani} 
\affil{Dipartimento di Fisica, Universit\'a di Pisa, Piazza Torricelli 2, I 56100 Pisa and INFN, via Livornese 1291, I 56010 Pisa, Italy - vittorio@astr18pi.difi.unipi.it}
 \author{S. Degl'Innocenti} 
\affil{ Dipartimento di Fisica, Universit\'a di Pisa, Piazza Torricelli 2, I 56100 Pisa
 and INFN, via Livornese 1291, I 56010 Pisa, Italy - scilla@astr18pi.difi.unipi.it}
\author{M. Marconi} 
\affil{Osservatorio Astronomico di Capodimonte, Via Moiarello 16, I 80131 Napoli, Italy - marcella@na.astro.it}

\begin{abstract}
We present a theoretical approach to infer information about RR Lyrae
variables from the morphology of their light curves. The method,
already successfully applied to the field first overtone variable U
Comae, is now tested on a RR$_{\mathrm ab}$ variable in the globular
cluster $\omega$~Cen. We show that, with this method, it could be
possible to give an estimate of the distance modulus of the cluster
with an uncertainty not larger than $\pm$0.1 mag. The predicted
variable luminosity and mass are well within the range of theoretical
evolutionary expectations giving the ``feeling'' of a full
compatibility between pulsational and evolutionary theory. However,
before completely relying on the method, further tests are needed; the
goal is that a well tested and well calibrated pulsational theory
could provide reliable distance moduli from just one (or few) RR Lyrae
stars.
\end{abstract}

\begin{figure}[h]
\plottwo{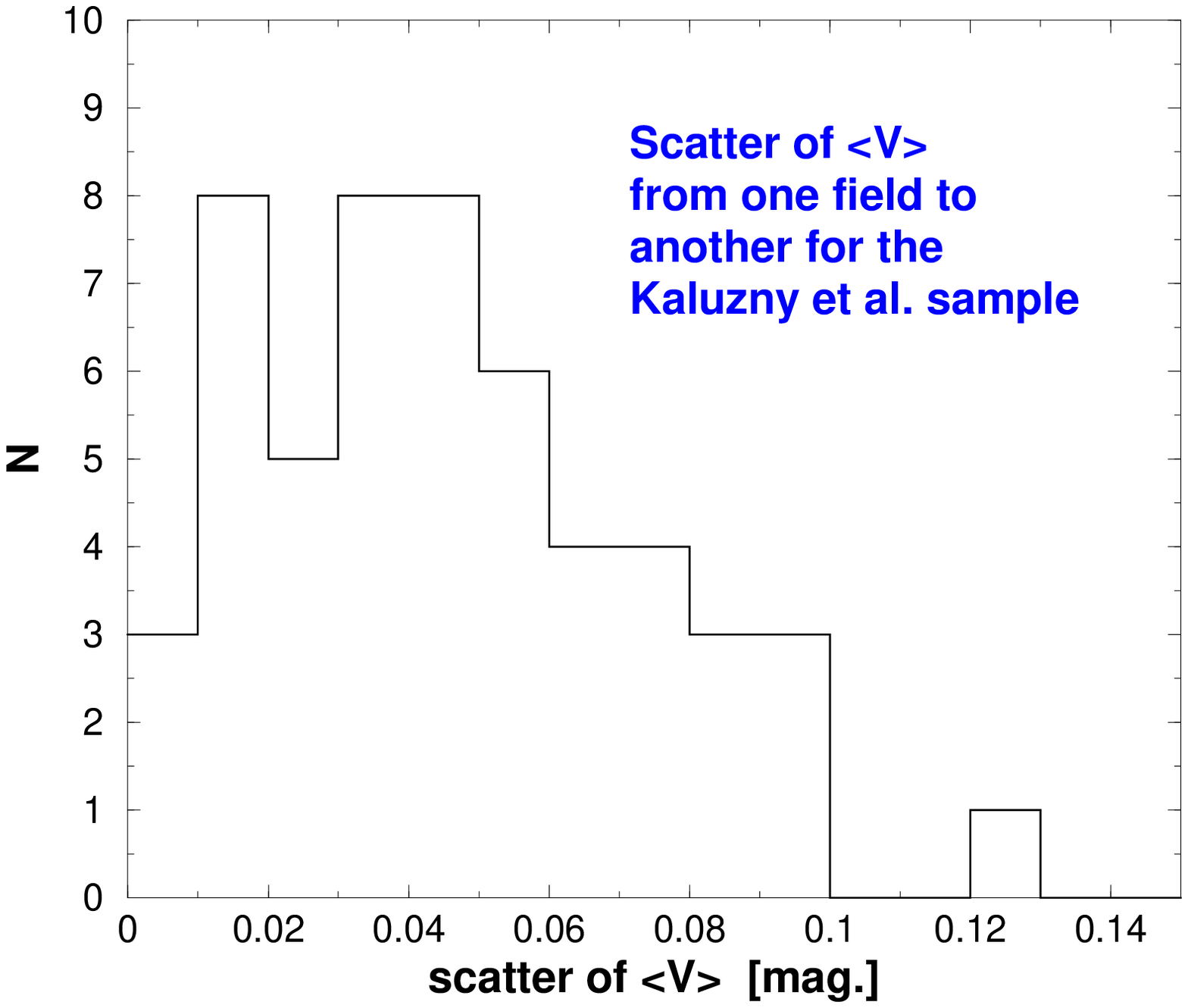}{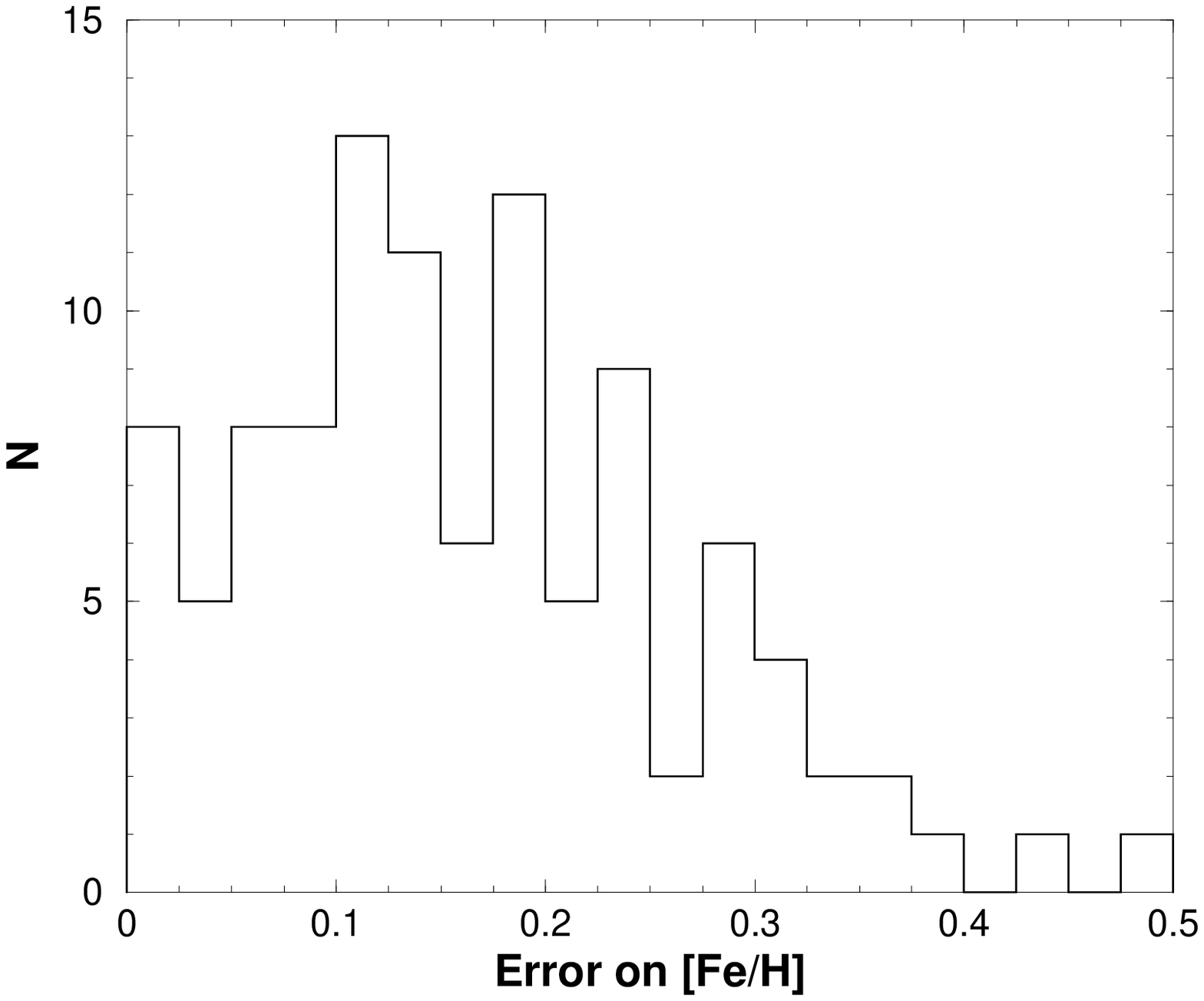}
\caption{Left panel: Frequency histogram for field to field scatter of RR Lyrae stars
mean visual magnitudes from Kaluzny et al. (1997). Right panel: Frequency histogram
of the nominal uncertainties in the metallicity estimates by Rey et al. (2000).}
\label{errori}
\end{figure}

\begin{figure}[h]
\vspace{-2cm}
\plotfiddle{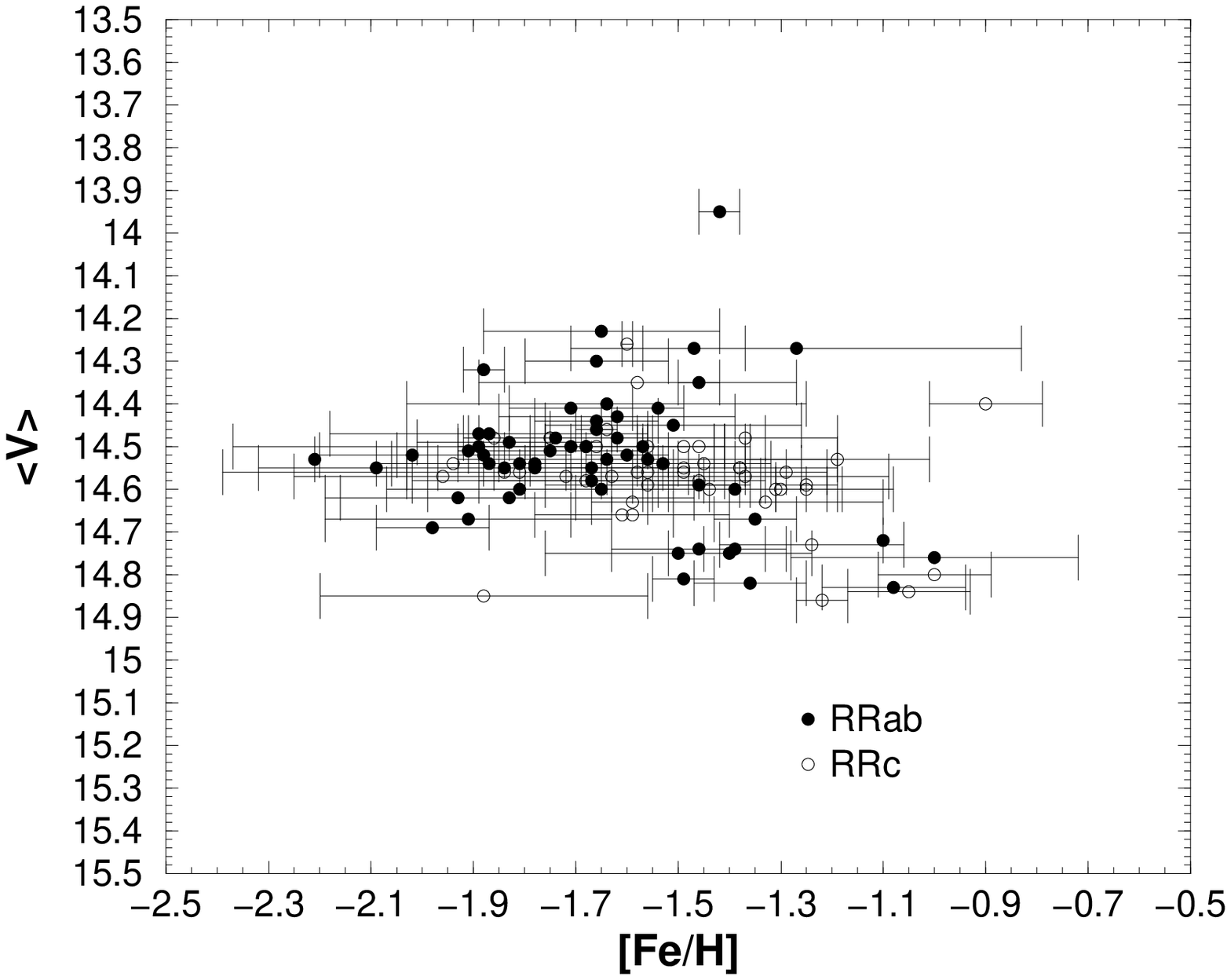}{8cm}{0}{50}{50}{-150}{-50}
\vspace{1cm}
\caption{The mean visual magnitude as a function of the metallicity
for the subsample of 105 RR Lyrae stars for which both mean magnitudes 
(Kaluzny et al. 1997) and metallicities (Rey et al. 2000) are available.}
\label{VFeH}
\end{figure}

\section{Introduction}
As well known, $\omega$~Cen is not only the largest globular cluster in
our Galaxy but also contains a huge amount of RR Lyrae variables,
of which more than 150 have already been detected in the cluster
region. According to these features, $\omega$~Cen could appear as
an excellent target for investigating the evolutionary status of
cluster stars and, in particular, to link the pulsational properties
of RR Lyrae variables to their evolutionary
parameters. Moreover, one can take advantage of quite a rich set of
observational data as the beautiful light curves provided by Kaluzny
et al. (1997) for 131 RR Lyrae stars, within the frame of the OGLE
experiment.
\begin{figure}[h]
\plottwo{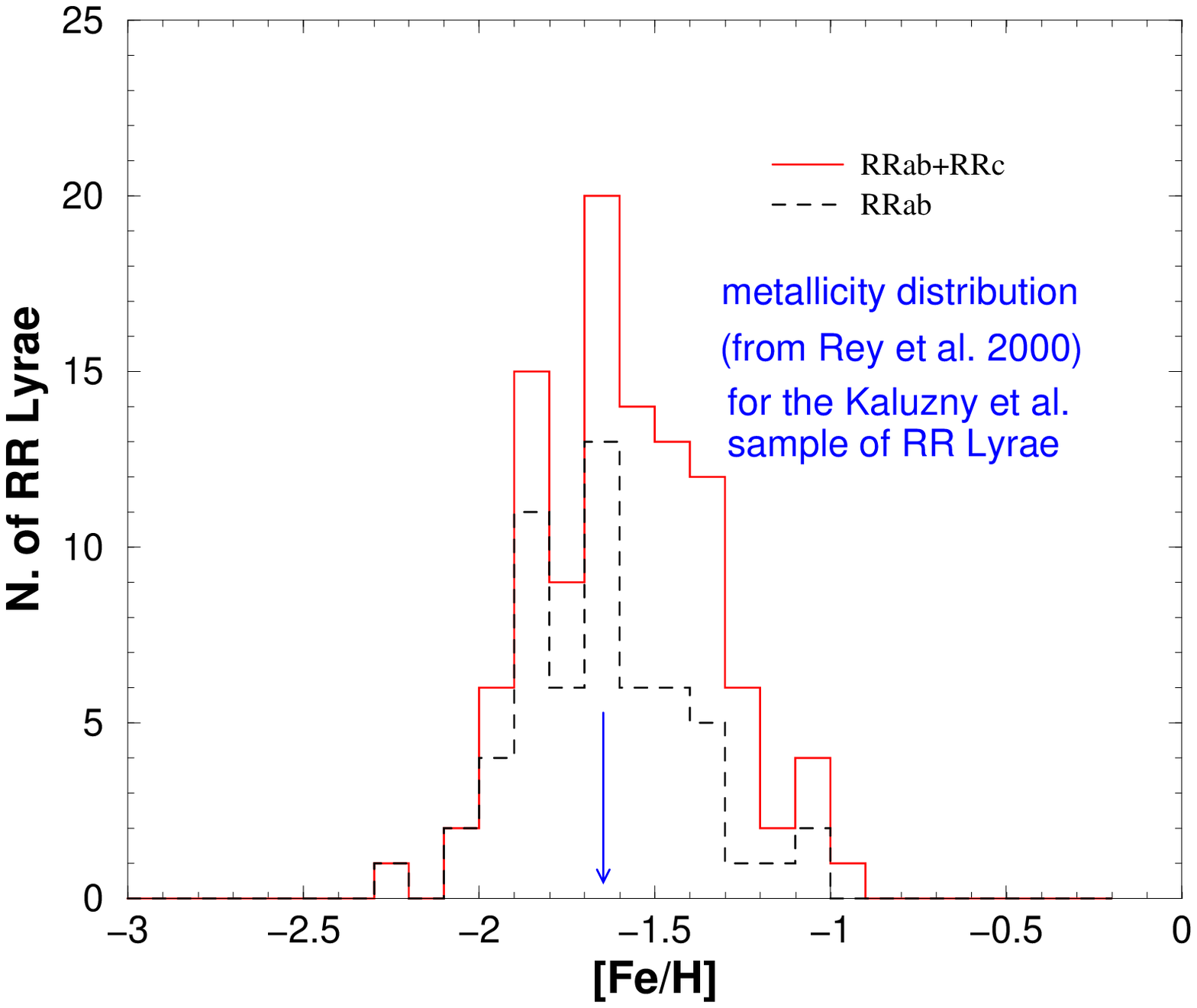}{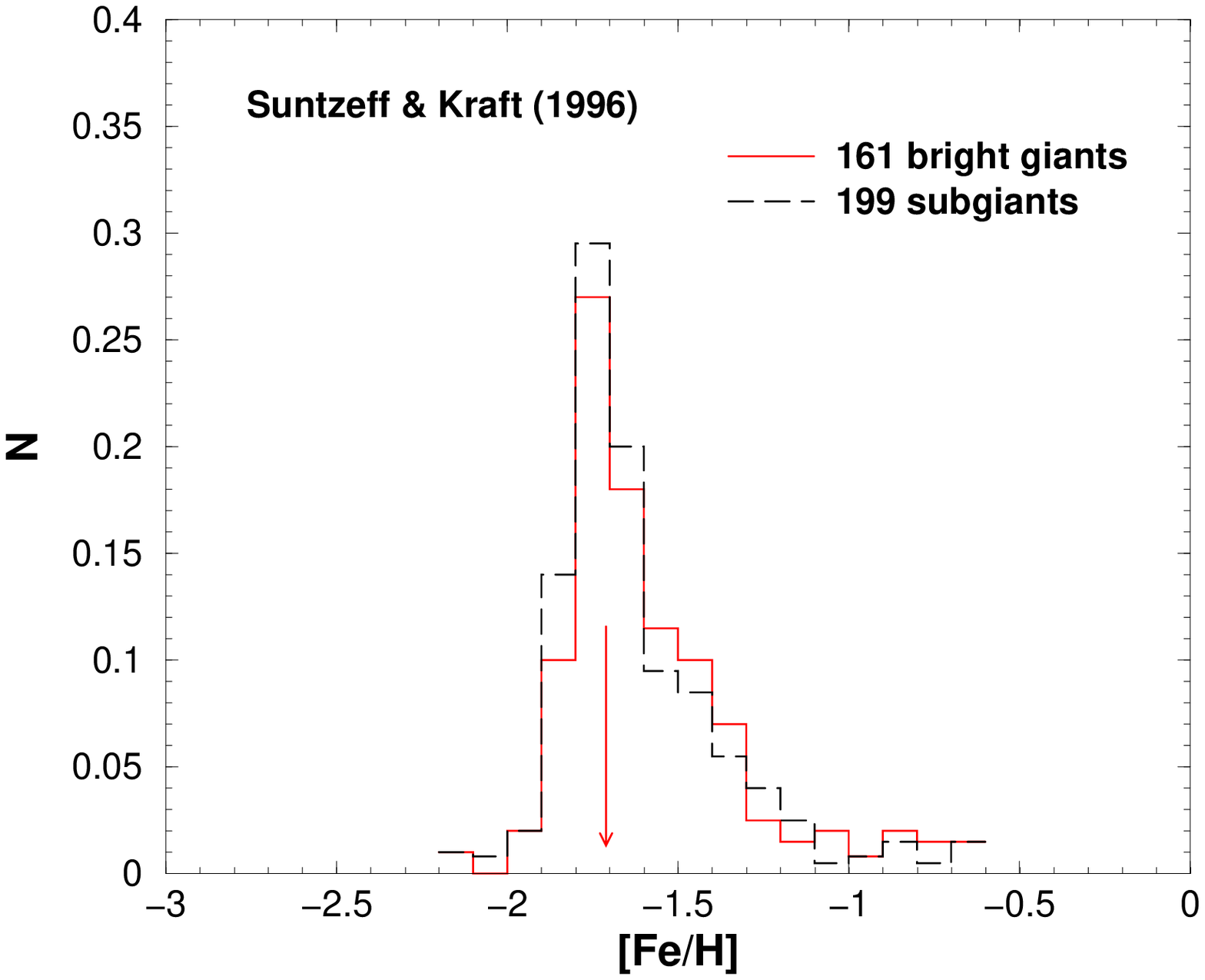}
\caption{Left panel: Metallicity distribution, from Rey et al. (2000),
for the Kaluzny et al. (1997) sample of RR Lyrae variables. The arrow indicates
the position of the peak distribution. Right panel: Metallicity
distribution, from Suntzeff \& Kraft (1996) for giants and subgiant
stars in $\omega$~Cen.}
\label{istoFe}
\end{figure}

\begin{figure}[h]
\vspace{-2cm}
\plotfiddle{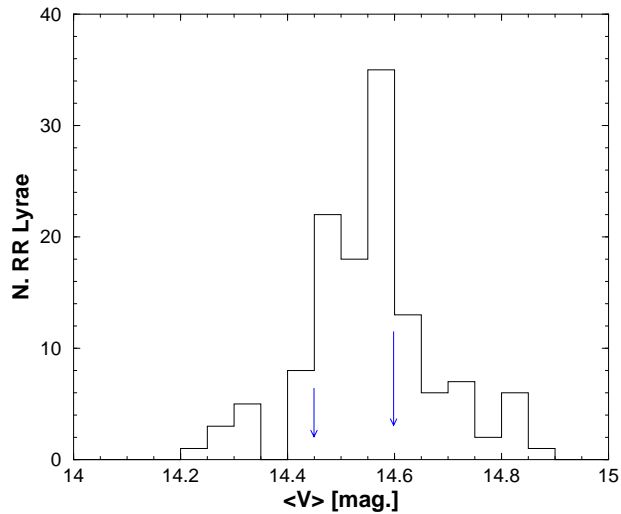}{8cm}{0}{50}{50}{-150}{-50}
\vspace{1cm}
\caption{Mean visual magnitude distribution for the RR Lyrae stars in the
Kaluzny et al. (1997) sample for which the metallicity estimate from Rey et
al. (2000) is available. The arrows indicate the range of visual magnitude
covered by the bulk or RR Lyrae variables.}
\label{NV}
\end{figure}

\begin{figure}[h]
\plottwo{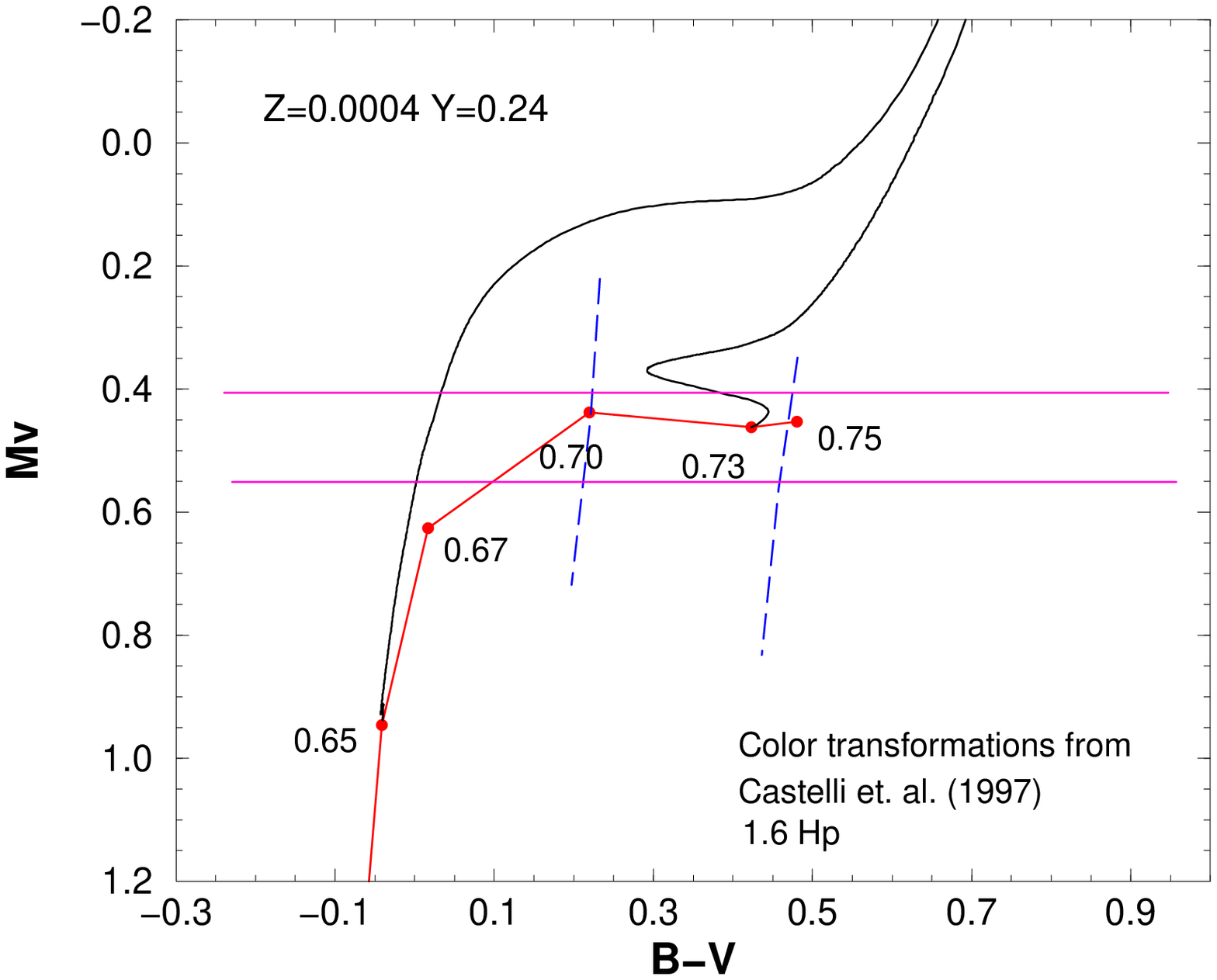}{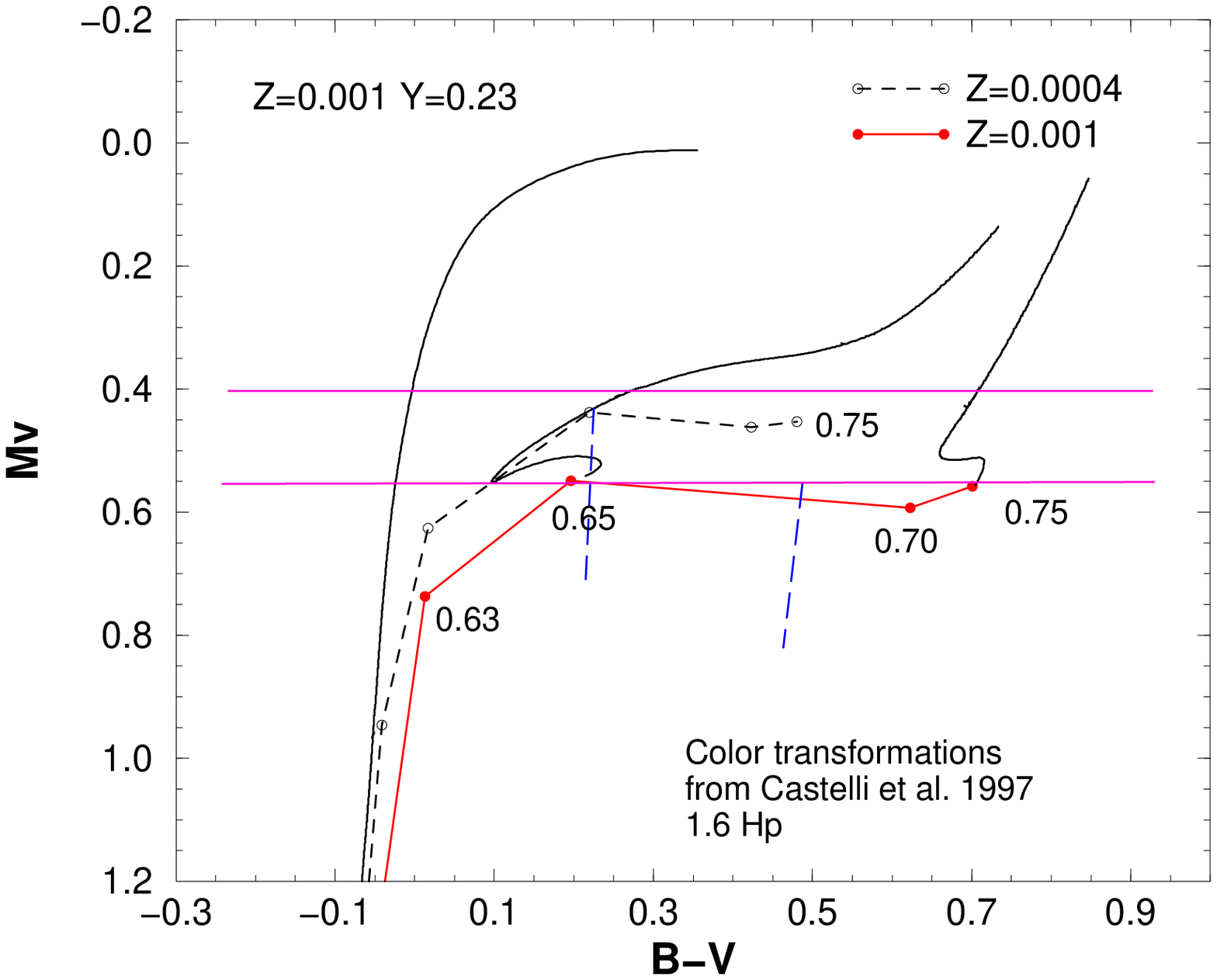}
\caption{Left panel: Zero Age Horizontal Branch (ZAHB) position for
star of the labelled masses and Z=0.0004 from present computations.
The central He burning evolution of selected masses is also shown. The
horizontal lines indicate the range of absolute visual magnitude for
the bulk of $\omega$~Cen RR Lyrae stars (see text). The vertical
dashed lines indicate the boundaries of the instability strip. Right
panel: the same as the in left panel but for a metallicity Z=0.001. The
ZAHB position of Z=0.0004 RR Lyrae variables is also shown for
comparison. Evolutionary tracks from Cassisi et al. (1998).}
\label{ZAHB}
\end{figure}
\begin{figure}[h]
\plotone{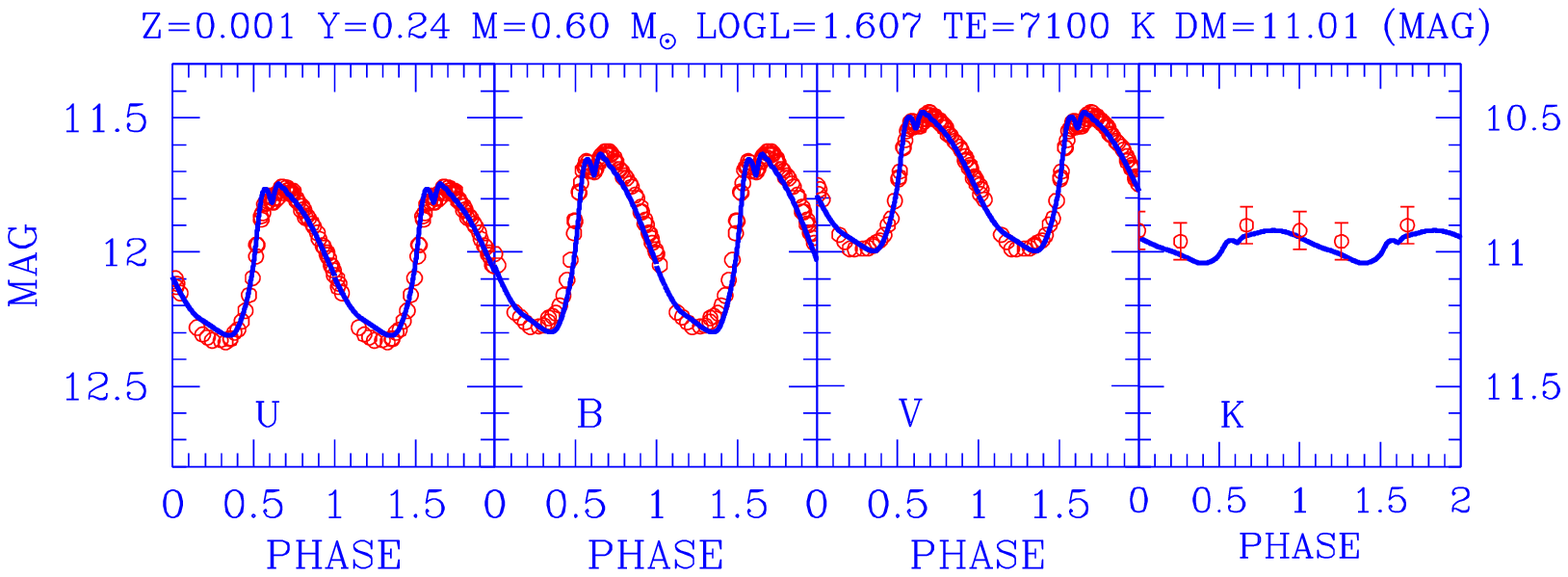}
\caption{Comparison between theory (lines) and observation (circles)
for the first overtone variable U Comae, as taken from Bono, Castellani \& Marconi (2000). From
left to right the panels refer to photometric data in U,B,V (Heiser
A.M. 1996) and K (Fernely, Skillen \& Burki 1993).}
\label{striscia}
\end{figure}

However, $\omega$~Cen definitely discloses the rare peculiarity of
member stars with a noticeable spread of metallicity. In probable
connection with such a feature, one also finds a large spread of RR
luminosities and a rather peculiar distribution of RR periods,
casting some doubts on the classification of this cluster as so a
``bona fide" Oosterhoff II. The occurrence of this metallicity spread
obviously complicates things, but it offers the exciting opportunity
of studying the metallicity effects on the luminosity of Horizontal
Branch stars, which is a long debated question connected with relevant
issues as globular cluster distances and ages.

Our attention on the problem was raised by the recent paper by Rey et
al. (2000), who presented metallicity estimates for 131 RR Lyrae variables in
$\omega$~Cen, as based on the use of the {\em hk} index. Some general
features of the cluster RR Lyrae population have been already
discussed in that paper, as well as in a following paper by Clement
\& Rowe (2000). In this context, we approached the problem on the
theoretical side, aiming to explore the degree of concordance between
the quoted sets of relevant observational data and the predictions of
current theories concerning both stellar evolution and stellar
pulsation. However, when entering into this matter one finds that
the available estimates for both the RR luminosities and metallicities,
though representing an highly valued step in our knowledge of cluster
stars, still leave some room for uncertainties. As a matter of the
fact, by comparing field to field mean magnitudes of RR Lyrae stars, as
given in Figure 1 (left panel) from the paper by Kaluzny et al. (1997), 
one finds that discrepancies can reach $\approx$0.1 mag.  Thus
one is dealing with rather accurate light curves but with a not
negligible uncertainty in the magnitude zero point.

The right panel of the same Figure 1 shows that a not negligible
uncertainty affects also metallicity estimates. Bearing in mind as a
warning such an ``imperfect" observational scenario, in the following
we will discuss observational data to the light of current theoretical
predictions.

\section{Luminosities of RR Lyrae variables}

By merging the already quoted data by Kaluzny et al. (1997) and by Rey et al. (2000), 
one is dealing with a subsample of 105 cluster RR Lyrae variables
with observational estimates for both mean magnitudes and
metallicity. Figure 2 shows the run of mean magnitudes versus
metallicity, where present uncertainties on the metallicity are
reported at their face values. The possible correlation between the
two quoted evolutionary parameters has been already addressed by Rey
et al. (2000) and it does not deserve further discussions. Here we
only notice that the rather large uncertainties do not allow a
significant comparison between theory and single stars, so that one
is forced to approach the problem along a``softer" route.

With this aim, Figure 3 (left panel) shows the metallicity frequency
histogram for RR Lyrae stars in the quoted subsample, as compared with
metallicity estimates as given for 161 bright giants and 199 subgiants
in the cluster by Suntzeff \& Kraft (1996: right panel). As already
discussed by Rey et al. (2000), when uncertainties in the calibration
of the {\em hk} index are taken into account, the two distributions
appear in reasonable agreement and, from the more accurate data on
giants and subgiants, one can safely assume that the large majority of
RR Lyrae stars should have a metallicity roughly in the range [Fe/H]$\sim
~-1.8\div~-1.4$.  From the magnitude frequency histogram of the
subsample given in Figure 4, one further obtains that the bulk of $\omega$~Cen
 RR Lyrae variables has a mean visual magnitude in the range V=14.45$\div$14.60 mag.
Now we are in the position of asking whether current evolutionary
theories are consistent with these observational data. This is of
course quite a limited question, but also (possibly) the only
significant question compatible with the scarce accuracy of the data
one is dealing with.

By adopting from Thompson et al. (2001) a cluster distance modulus
(V-M$_{\mathrm V}$)= 14.05 $\pm$ 0.11, the observational scenario tell us that RR
Lyrae stars, approximately covering the metallicity range $0.0002 \leq Z \leq
0.002$, span a range of luminosity between M$_{\mathrm V}$= 0.40 and 0.55, with the
additional uncertainty of $\pm 0.11$ mag., due to the uncertainty in the
distance modulus.  Figure 5 compares this observational evidence with
theoretical predictions concerning the luminosity of Horizontal Branch
stars with Z=0.0004 and Z=0.001, as given by Cassisi et al. (1998) and
by present computations. One derives the comforting evidence that the
observed spread in luminosity appears in good agreement with 
current theoretical predictions.  Formally, one  also finds a good
agreement regarding the absolute magnitude of the stars. However,
owing to the quoted indetermination in the cluster distance modulus, such a
nominal agreement cannot be used to disentangle the thorny problem of
theoretical HB luminosities, which is still waiting for a firm
settlement (see, e.g., Caputo et al. 1999; Castellani et al. 2000, for
a general discussion on that matter) and which would require much
firmer observational constraints.

\section{The pulsational route}


\begin{figure}[h]
\plottwo{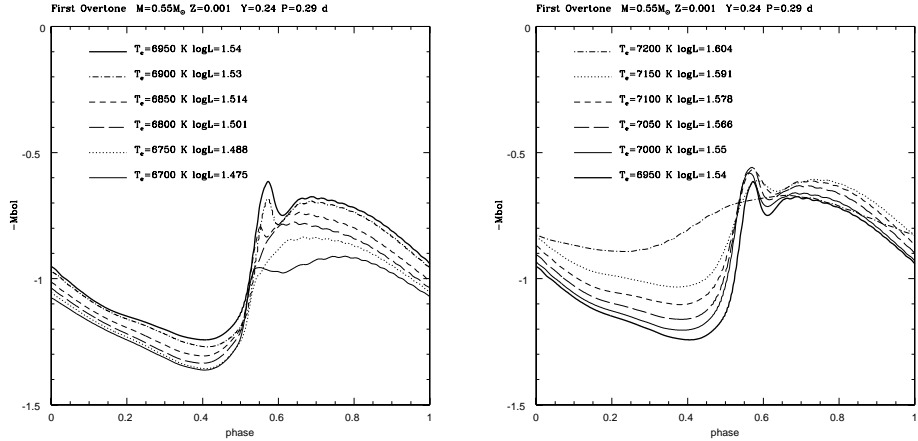}{castellani_fig7b.epsi}
\caption{Theoretical light curves for  first overtone pulsators with
fixed period (0.29 days) and mass (0.55 M$_{\odot}$) when the temperature
is varied by steps of 50$^{\mathrm o}$K. Chemical composition as labelled.}
\label{calde}
\end{figure}


In the previous section we dealt with observations concerning the CM
diagram location of the $\omega$~Cen RR Lyrae variables, without any
regard to the observed pulsational properties. However, in the last
years a great deal of investigation has been devoted to the connection
between pulsational properties of RR Lyrae variables and structural
parameters, such as the pulsator mass, luminosity and effective
temperature. As well known, a firm knowledge of the mechanisms
governing such a connection would be of paramount relevance, allowing
the use of easily observed pulsational features to obtain independent
information on the evolutionary status of the pulsating structure. As
a matter of fact, the use of pulsational periods to constrain
evolutionary stellar structures has already a long history, starting
from the pioneering paper by van Albada \& Baker (1971).  However,
only more recently the development of hydrodynamic non linear
computational codes has allowed the prediction of pulsational
amplitudes, adding further constraints to the pulsator structures (see,
e.g., Bono \& Stellingwerf 1994; Bono et al. 1997a). At the same time,
one has to notice that periods and amplitudes are only a partial and
strongly limited parameterization of the star pulsational behavior,
whereas modern computations do produce predictions about the whole
pulsator light curve, thus giving much more severe constraints for the
comparison with observation. In this context, the use of Fourier
parameters (see, e.g., Kovacs \& Jurcsik 1996; Jurcsik 1998; Poretti
1999) is indeed a new way to take into account the shape of the light
curve in the discussion of pulsator properties. However, the best fit
of the whole light curve appears as an obvious ultimate goal of
pulsational theories. We have recently followed this way, attempting
to reproduce the peculiar light curve of the field RR$_{\mathrm c}$ U
Comae (Bono, Castellani \& Marconi 2000).
 
The adopted procedure was based on the well known occurrence that
pulsators periods are expected to depend on the star mass, luminosity
and effective temperature.  According to such an occurrence, given the
observed period (P=0.29 days) and by assuming the pulsator mass
in the range predicted by evolutionary theories, for each assumed
value of the luminosity L, there is only one value of the effective
temperature T$_e$. Thus one has to investigate whether in a suitable
range of luminosity can exist a theoretical light curve reproducing
the observed one. This approach was indeed quite successful and
Figure 6 shows that theoretical predictions reach an excellent
agreement with observational data in various photometric bands. 
The fit is obtained for a quite reasonable value of the predicted
luminosity, also in agreement with previous independent estimates of
such a parameter.

To appreciate the sensitivity of the method let us show in Figure 7
the variation of the predicted light curve, at fixed period and
pulsator mass, when the effective temperature is varied by steps of
only 50$^{\mathrm o}$K (and M$_{\mathrm bol}$ by steps of $\sim$0.03
mag. to keep the period constant). Taking the theoretical results at
their face value one should conclude that the fitting of the light
curve should be able to fix the star luminosity and effective
temperature with an unprecedented degree of accuracy!

According to this successful result, in the occasion of this meeting
we were stimulated to explore whether a similar approach can be
repeated to derive information on the luminosity of $\omega$~Cen
RR variables. In this case we choose an ab-type variable (99B in the
Kaluzny list) whose light curve is reported in Figure 9, with the
labelled value of the observed period and mean V-magnitude. The
metallicity estimate for this star (from Rey et al. 2000) is [Fe/H] =
-1.74$\pm$0.05; thus all computations have been performed under the
assumption Z=0.0004, even though the results are barely sensitive to
the metallicity value within the metal poor range (Bono et al. 1997b,
Caputo, Marconi, Santolamazza 1998).

However, in this case we will adopt a slightly different approach which
avoid assumptions about the star masses and  which runs as follows:

i) Let us again assume the star luminosity L as a free parameter. 

ii) Given the observed period, for each adopted value of the luminosity
and for each assumed effective temperature, there is only one value for
the mass and thus only one predicted light curve.

iii) As shown in Figure 8 (left panel), theory predicts that at each given luminosity
level the amplitude is continuously decreasing with the effective temperature.

iv) By requiring that the predicted light curve shows the observed
amplitude, for each assumed luminosity level, one finds one and only
one predicted light curve.

As a result, for each assumption about the star luminosity, one
obtains a value of T$_{e}$, a value for the stellar mass, a light
curve with the observed period and amplitude and a mean absolute
visual magnitude for the pulsator, i.e., a cluster distance
modulus. One eventually is dealing with a sequence of light curves
only depending on the assumed star luminosity, and the best fitting
with observation -if any- has to be found among this linear set of
results.

The procedure is made easier by the (theoretical) evidence that for
each given effective temperature the pulsational amplitude is only
slowly depending on the assumed luminosity level, as shown in Figure 8
(right panel). Thus the effective temperature derived when exploring
the first luminosity level gives a useful indication of the
temperatures needed at different luminosities to account for the
observed pulsational amplitude.

As a result of this procedure we find that a satisfactory, though not
perfect, best fitting requires a star luminosity of
logL/L$_{\odot}$$\approx$1.77, surprisingly corresponding to the cluster
distance modulus 14.05 already adopted in the previous section (see
Figure 9, upper right panel).  As shown in the same Figure 9 (lower panels),
by moving the adopted distance modulus by $\pm$ 0.15 mag. the fitting
appears definitely worse, so that one can estimate a (nominal)
uncertainty in the procedure not larger than $\pm$ 0.1 mag, i.e., a
bit larger than in the quoted U Comae case but still reasonably small.

Let us finally notice that the fitting has been achieved for a mass
value (M=0.75M$_{\odot}$) which appears well within the range of
theoretical evolutionary expectations, giving to the whole predicted
scenario at least the ``flavour" of a full compatibility.

\section{Final remarks}

In the previous section we have shown how modern pulsational
computations, when applied to reproduce the observed light curves of
variable stars, appear in principle able to give rather firm
constraints on the evolutionary parameters of the pulsating structures
and, in turn, on the distance modulus of the parent cluster, if
any. However, one cannot forget that this is just a ``theoretical
truth" and the question arises of how true is this truth!

As a matter of fact, one has to remind that modern pulsational
hydrodynamical computations are quite sophisticated, requiring a
detailed mathematical description of the turbulent convective
model and of the coupling between dynamical and convective
motions. As a consequence, one cannot exclude that theoretical
results are affected by some unknown systematic errors. Thus, to tell
the truth, we were already rather surprised of the best-fitting
capability of the adopted pulsational theory, as well as of the quite
reasonable scenario emerging from this fitting. In our feeling such an
agreement cannot be casual, indicating that modern pulsational
computations are, at least, on the right way.

However, deep tests of the theory are still needed. As an example, it
would be very useful to have much firmer information on the pulsator
effective temperature, to be compared with the temperature required by
the pulsational fitting. The availability of observational velocity
curves will be also of great relevance, allowing a tight comparison
with the predicted pulsational behaviour (see also Bono,
Castellani \& Marconi 2000). Last, but not least, precise [Fe/H] value
would be highly desirable, since some details of the predicted light
curves could depend on the adopted metallicity.

Similar tests of the pulsational theory would be of paramount
relevance, since - as a final goal - a well tested and well calibrated
pulsational theory will provide reliable distance moduli from just one
(or few) RR Lyrae variables, supporting and improving the theoretical scenario
discussed by Wood (1997) in connection with bump Chepheids in the
Large Magellanic Cloud.

\vspace{3cm}
\begin{figure}[h]
\plottwo{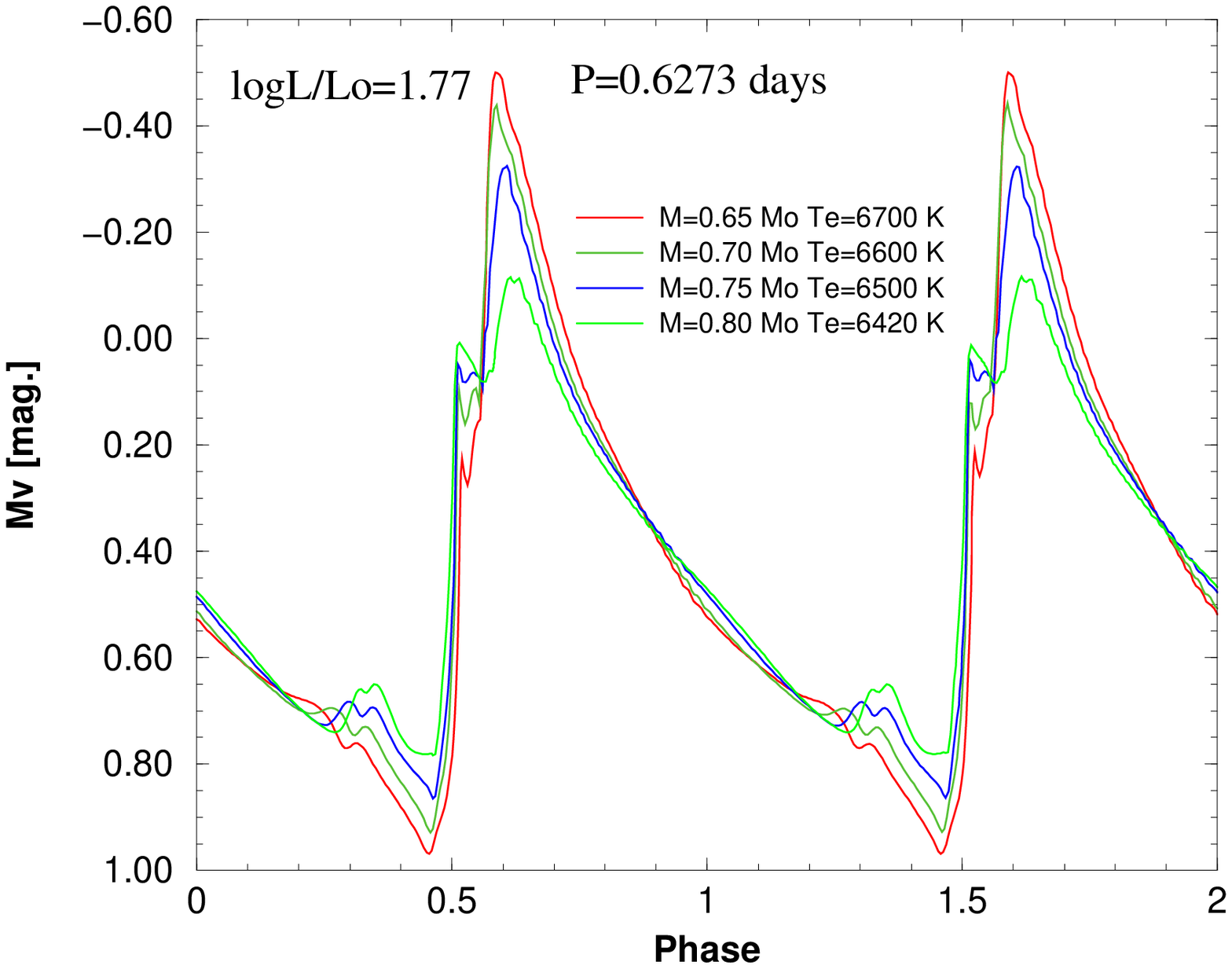}{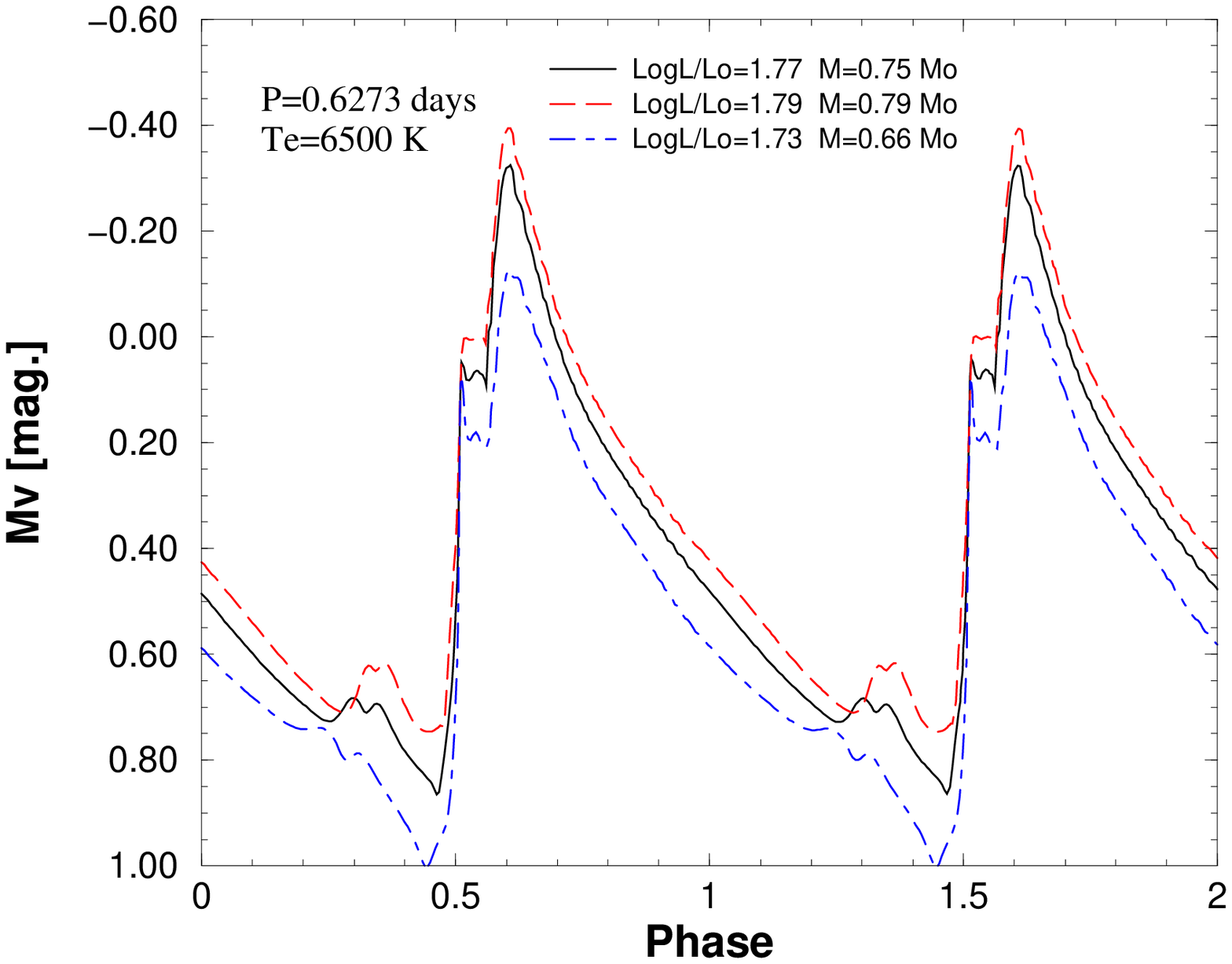}
\caption{Left panel: theoretical light curves at fixed period (0.627 days) and
luminosity level (logL/L$_{\odot}$=1.77) for different effective
temperatures (and masses), as labelled. Right panel: theoretical light
curves at fixed period (0.627 days) and effective temperature (6500$^{\mathrm o}$K)
 for three different luminosity levels (and masses), as labelled.}
\label{curva99te}
\end{figure}


\begin{figure}[h]
\plotone{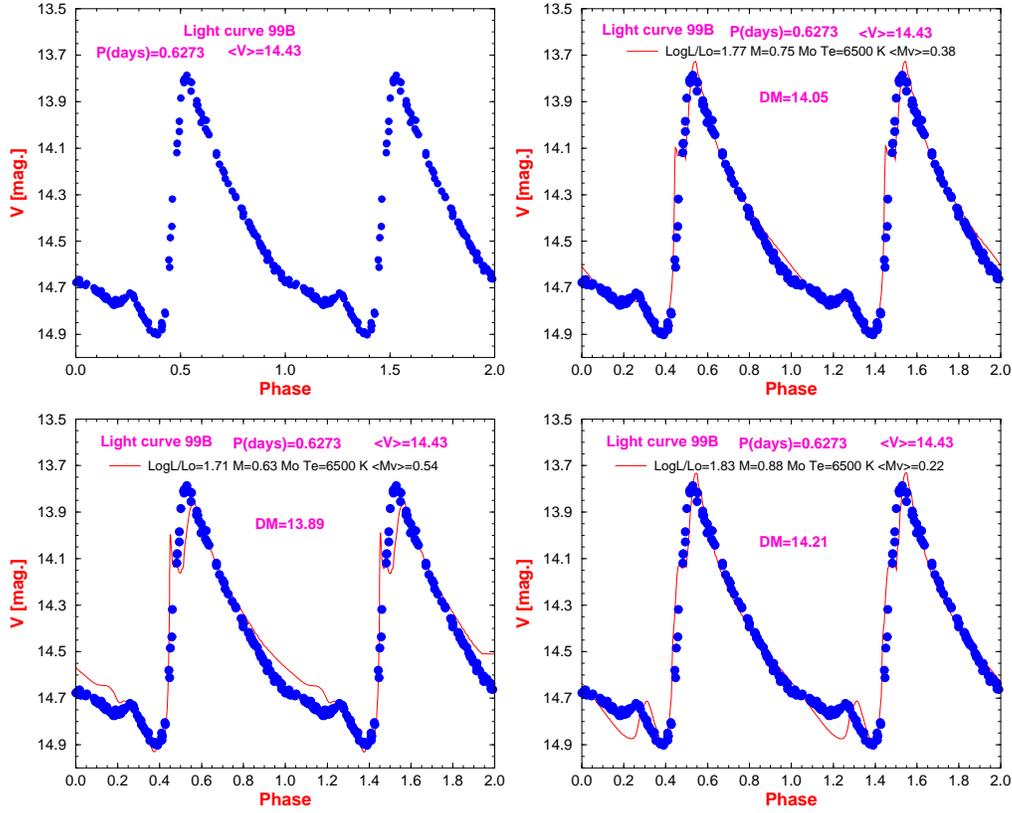}
\caption{Upper left panel: observational light curve for the
RR$_{\mathrm ab}$ variable 99B in the Kaluzny et al. (1997)
sample. Upper righ panel: best fit of the observational light curve
with a theoretical light curve for the labelled values of luminosity,
effective temperature and mass. Lower panels: as in the upper right
panel but when the luminosity level of the theoretical light curves is
varied by $\pm$ 0.15 mag with respect to the best fit. The stellar
luminosity, temperature and mass are labelled. The distance modulus
adopted for the fits is also shown.}
\label{curva99}
\end{figure}

\vspace{1cm}
\acknowledgements

We warmly thank Filippina Caputo for a critical reading of the manuscript.


\end{document}